\documentstyle[11pt,a4,epsfig,amssymb]{article}

\makeatletter
\@addtoreset{equation}{section}
\makeatother

\newcommand{\comment}[1]{{}}

\newcommand{\beq}{\begin{equation}}
\newcommand{\eeq}{\end{equation}}
\newcommand{\ba}[1]{\begin{array}{#1}}
\newcommand{\ea}{\end{array}}
\newcommand{\bea}{\begin{eqnarray}}
\newcommand{\eea}{\end{eqnarray}}

\newcommand{\sz}{\scriptsize}

\newcommand{\jump}[1]{$^{\mbox{\sz {#1}}}$\/}

\title{
\begin{flushright} {\normalsize \today} \end{flushright}
\vspace{6ex}
{\bf
Top-Higgs Yukawa Coupling Measurement at a Linear $e^+e^-$ Collider
} \\
\vspace{1ex}
}

\date{}

\author{
\large {\sc Aurelio Juste\jump{1}\/, Gonzalo Merino\jump{2}\/} \\
\\
\normalsize  1) Fermi National Accelerator Laboratory, \\
\normalsize  P.O. Box 500, MS 357, \\
\normalsize  Batavia, IL 60510, \\
\normalsize  Phone: 1\,(630)\,840\,-6565 \ \ Fax: 1\,(630)\,840\,-8481 \\
\normalsize  e-mail: juste@fnal.gov \\
\\
\normalsize  2) Institut de F\'{\i}sica d'Altes Energies, \\
\normalsize  Universitat Aut\`{o}noma de Barcelona, \\
\normalsize  E-08193 Bellaterra (Barcelona), Spain. \\
\normalsize  Phone: 34\,(93)\,581\,-2846\ \ Fax: 34\,(93)\,581\,-1938 \\
\normalsize  e-mail: merino@ifae.es \\
}

\begin{document}

\maketitle
\begin{abstract}
A feasibility study  of the measurement of the top-Higgs Yukawa coupling
at a future linear $e^+e^-$ collider operating at $\sqrt{s}=800$ GeV is
presented. As compared to previous existing studies, much effort has been
put in a ``realistic simulation'' by including irreducible+reducible
backgrounds, realistic detector effects and reconstruction procedures and
finally, a multivariate analysis. Both hadronic and semileptonic decay
channels have been considered.
\end{abstract}

\newpage

\section{Introduction}
The theory of electroweak interactions has been succesfully tested so far to an
extremely high degree of accuracy. However, one of its key elements, the Higgs mechanism,
remains to be tested experimentally.
It is through the interaction with the ground state Higgs field
that the fundamental particles acquire mass, which in turn sets the scale
of the coupling with the Higgs boson. Once the Higgs boson is found (if ever),
all its properties have to be accurately measured: mass, width, etc, and indeed
its couplings to bosons and fermions. The couplings to the $Z$ and $W$ gauge bosons
can be measured from the Higgstrahlung process: $Z\to ZH$~\cite{higgstrahlung} 
and the fusion processes: $WW,\: ZZ \to H$~\cite{fusion}.
On the other hand, the top quark provides a unique opportunity to measure 
the Higgs Yukawa coupling to fermions. Being proportional to the fermion mass,
\begin{eqnarray}
g_{ffH} = \frac{m_f}{v},
\end{eqnarray}
\noindent with $v = (\sqrt{2} \: G_F)^{-1/2} \simeq 246$ GeV, the top-Higgs 
Yukawa coupling is the largest among the different fermions:
$g_{ttH}^2 \simeq 0.5$ to be compared for instance with
$g_{bbH}^2 \simeq 4 \times 10^{-4}$.

The process $e^+e^- \to t\bar{t}H$ provides a chance for a direct 
measurement of the top-Higgs Yukawa coupling~\cite{tthlo1,tthlo2} in the ``Light Higgs
Scenario'' ($M_H \leq 2 m_t$), which seems to be favored by the present
precision electroweak data ($M_H = 76^{+85}_{-47}$ GeV as reported in ~\cite{lepewwg}).
On the other hand, the current limit form direct search at LEP2 is $M_H > 95.2$ GeV at 95\% CL~\cite{limit}.
The total cross-section depends sensitively on the top-Higgs Yukawa coupling, which can
thus be inferred from the comparison of the measured total cross-section with the
theoretical expectation as a function of $g_{ttH}$.
This measurement of $g_{ttH}$ is direct as compared to the
indirect determination via its effect in the interquark potential near the
$t\bar{t}$ production threshold, which affects some threshold observables~\cite{threshold}.

In this study we are going to assume that the MSM Higgs boson has already been discovered
and its mass measured to be $M_H = 120$ GeV. For $M_H = 120$ GeV, 
the Higgs decays dominantly to $b\bar{b}$ (BR($H\to b\bar{b})\simeq 77\%$),
and assuming BR($t\to W b) = 100\%$, this
leads to multi-jet event topologies involving 4 b-jets in the final state.
Therefore, one of the crucial experimental aspects will be flavor tagging.

Previous studies on the feasibility of this measurement have already been 
performed~\cite{previous}, but they were assuming a too simplistic simulation of detector
effects (in particular b-tagging, which is critical here) and/or considering
too few background processes. More complete studies have recently been presented~\cite{recent}.

\section{Theoretical Scenario}
\begin{figure}[h]
\begin{center}
\mbox{
\epsfig{file=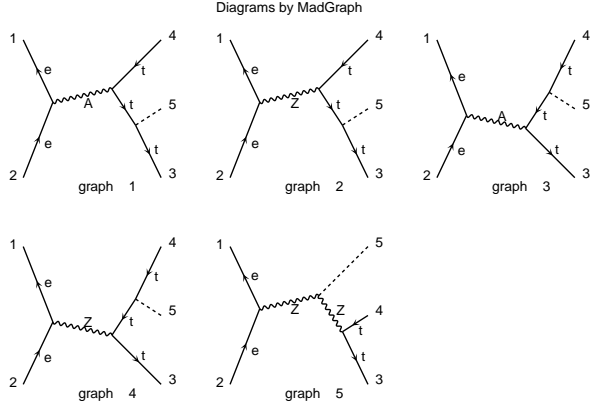,width=11cm}
}
\end{center}
\caption{\label{epem_ttbh}
\protect\footnotesize
Tree level diagrams diagrams contributing to the process $t\bar{t}H$.}
\end{figure}

\begin{figure}[h]
\begin{center}
\mbox{
\epsfig{file=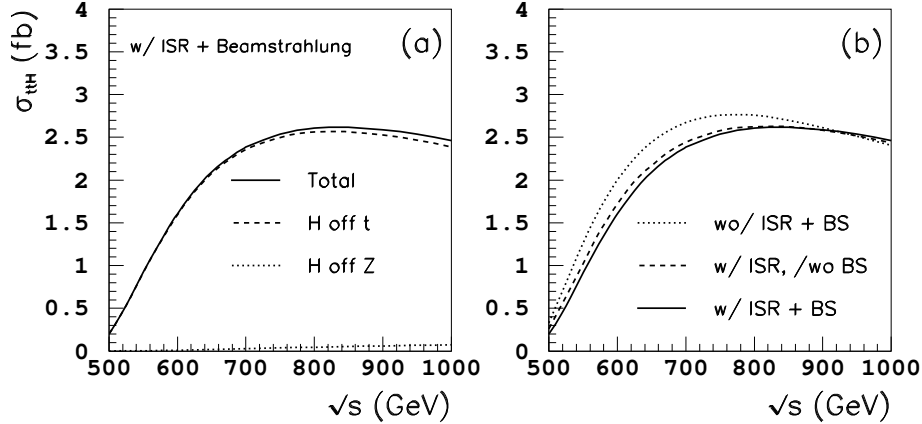,width=13cm}
}
\end{center}
\caption{\label{sigmatth}
\protect\footnotesize
Total cross-section for $t\bar{t}H$ at lowest order, for $m_t = 175$ GeV and
$M_H = 120$ GeV, as a function of the center-of-mass energy. In (a) the two
contributions: $H$ off the $t$ (dashed) and $H$ off the $Z$ (dotted) are
explicited. The effect of radiative processes in the initial state (initial
state radiation and beamstrahlung) on the total cross-section is illustrated in (b).}
\end{figure}

At lowest order 
there are 5 diagrams contributing to this process, as shown in Fig.~\ref{epem_ttbh}.
The dominant contribution comes from $\gamma$-exchange with 
the $H$ being radiated off the $t$ or the $\bar{t}$. The diagram in which the $H$ is
radiated off the $Z$ constitutes just a small correction,
so that the total cross-section is to a good approximation $\propto g_{ttH}^2$
(see Fig.~\ref{sigmatth}a).

As it can be observed in Fig.~\ref{sigmatth}a, the total cross-section
decreases at low $\sqrt{s}$ due to phase-space restrictions and
at high $\sqrt{s}$ due to unitarity.
Radiative effects in the initial state (initial state radiation and
beamstrahlung) become increasingly important for high $\sqrt{s}$ and
significantly distort the $t\bar{t}H$ lineshape, as shown in
Fig.~\ref{sigmatth}b. The main effect is to shift the maximum of the
cross-section towards higher $\sqrt{s}$, which for $M_H = 120$ GeV is
around $\sqrt{s} \simeq 800$ GeV. Initial state radiation turns out
to be the dominant radiative process in order to decrease the
sensitivity of the total cross-section on the Yukawa coupling.

Recently, the NLO QCD corrections to the total cross-section have
been computed~\cite{tthnlo}. They turn out to be important at moderate energies,
due to rescattering diagrams, which are generated by the Coulombic gluons 
exchange between the top quarks near the $t\bar{t}$ threshold.
As a consequence, the total cross-section
can be enhanced by a factor 2 with respect to LO whereas,
since virtual and soft-gluon radiation are the dominant corrections, the Higgs and top quark
energy and angular distributions are hardly changed.
However, above threshold these corrections to the total cross-section 
are small ($\sim 5\%$ at $\sqrt{s} = 1$ TeV) 
and negative, so for the purpose of this analysis they can be safely neglected.

Since the Yukawa coupling is determined from the cross-section measurement,
it is straightforward to estimate the expected statistical and some systematic
uncertainties on $g_{ttH}$ for a given selection with efficiency $\epsilon$ and purity 
$\rho$, applied on a data sample corresponding to an integrated luminosity $L$:
\begin{eqnarray}
 \biggl (\frac{\Delta g_{ttH}}{g_{ttH}} \biggr)_{stat} &\simeq& (\Delta g_{ttH}^2)_{stat} = 
\frac{1}{S_{stat}(g_{ttH}^2)\sqrt{\epsilon \rho L}}, \\
 \biggl (\frac{\Delta g_{ttH}}{g_{ttH}} \biggr)_{syst} &\simeq& (\Delta g_{ttH}^2)_{syst} = 
\frac{1}{S_{syst}(g_{ttH}^2)} \left [ 
\frac{1-\rho}{\rho}\frac{\Delta \sigma_B^{eff}}{\sigma_B^{eff}} \oplus
\frac{1}{\rho} \frac{\Delta L}{L} \oplus \frac{\Delta \epsilon}{\epsilon}\right ] , 
\label{gttherrors}
\end{eqnarray}
\noindent where $(\Delta g_{ttH}/g_{ttH})_{syst}$ accounts for the uncertainties
in the effective background cross-section (after selection), the integrated luminosity
and the selection signal efficiency. $S_{stat}(g_{ttH}^2)$ and 
$S_{syst}(g_{ttH}^2)$ are the so-called ``sensitivity factors'',
defined as:
\begin{eqnarray}
 S_{stat}(g_{ttH}^2) &=& \frac{1}{\sqrt{\sigma_{ttH}}} \Bigg\vert \frac{d\sigma_{ttH}}{dg_{ttH}^2} \Bigg\vert, \\
 S_{syst}(g_{ttH}^2) &=& \frac{1}{\sigma_{ttH}} \Bigg\vert \frac{d\sigma_{ttH}}{dg_{ttH}^2} \Bigg\vert,
\end{eqnarray}
\noindent which are a function of $\sqrt{s}$ for a fixed $M_H$. 

\begin{figure}[h]
\begin{center}
\mbox{
\epsfig{file=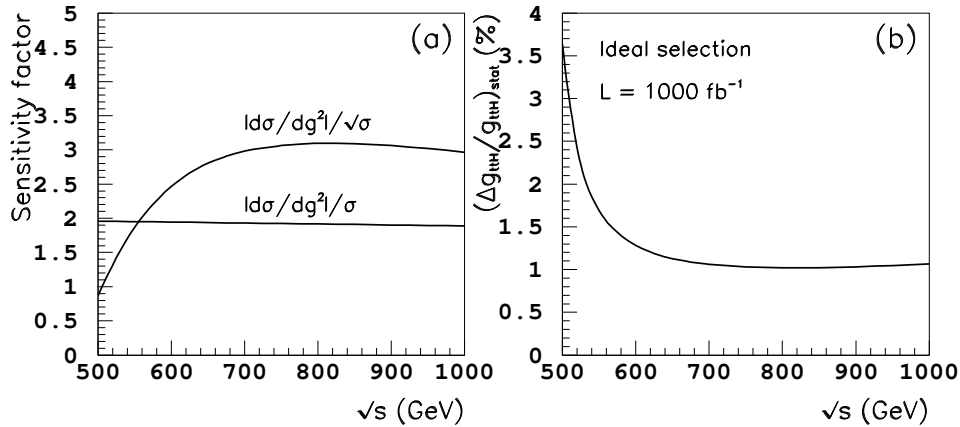,width=13cm}
}
\end{center}
\caption{\label{sens}
\protect\footnotesize
Sensitivity of the total cross-section on the top-Higgs Yukawa coupling as a function of the
center-of-mass energy for $M_H = 120$ GeV and including radiative effects:
(a) ``sensitivity factors'', (b) statistical uncertainty for an ideal selection assuming
an integrated luminosity of 1000 fb$^{-1}$.
}
\end{figure}

The sensitivity factors as a function of the center-of-mass energy are shown
in Fig.~\ref{sens}a for $M_H = 120$ GeV. As it can be observed, 
$S_{stat}$ reaches a ``plateau'' for
$\sqrt{s} \geq 700$ GeV, whereas $S_{syst}$ is essentially independent of $\sqrt{s}$.
At $\sqrt{s}=800$ GeV, the respective values are:
$S_{stat}\simeq 3.09$ fb$^{1/2}$ and $S_{syst}\simeq 1.92$.
Therefore, assuming an ideal selection ($\epsilon=100\%$ and
$\rho=100\%$), a statistical precision of around 1\% could be achieved in $g_{ttH}$
for $\sqrt{s} \geq 700$ GeV and $L = 1000$ fb$^{-1}$.
In a more realistic situation of $\epsilon=5\%$ and $\rho=50\%$, the statistical uncertainty
would be $\simeq 6.5\%$, whereas the systematic uncertainty would be
dominated by the uncertainty in the background normalization,
if one assumed that both the
signal selection efficiency and integrated luminosity can be known at the 1\% level or better.

\section{Simulation Aspects}
In order to simulate the signal and the $t\bar{t}Z$ background, we
have written a Monte Carlo event generator by using the squared matrix element
as computed by the program MADGRAPH~\cite{madgraph}
and the HELAS~\cite{helas} subroutines. The top and Higgs masses are assumed to be
$m_t = 175$ GeV and $M_H = 120$ GeV, respectively. The top width is
computed including NLO QCD corrections. Tops are generated off-shell by 
including the corresponding Breit-Wigner distributions in the differential
cross-section. 
The total cross-section and differential
distributions are found to be in good agreement with the calculation
in~\cite{tthlo1}.
The rest of backgrounds have been generated with PYTHIA~\cite{jetset}. 
Interferences between signal and backgrounds have been neglected.
The event samples have been generated
at $\sqrt{s} = 800$ GeV, including initial state radiation and
beamstrahlung. Initial state radiation has been considered in the structure
function approach and beamstrahlung has been generated with the aid of
the CIRCE program~\cite{circe}.
Fragmentation, hadronization and particle decays are handled
by JETSET~\cite{jetset}, with parameters tunned to LEP2 data.

\subsection{Detector Simulation}
Once the events have been generated, they are processed through
a fast simulation~\cite{simdet} of the response of a
detector for the TESLA linear collider. The detector components,
which are assumed to be:
\begin{itemize}
\item a vertex detector,
\item a tracker system with main tracker, forward tracker and
forward muon tracker,
\item an electromagnetic calorimeter,
\item a hadronic calorimeter and
\item a luminosity detector,
\end{itemize}
\noindent are implemented according to the TESLA Conceptual Design
Report~\cite{teslacdr}.

This fast detector simulation provides a flexible tool since its performance
characteristics can be varied within a wide range. The calorimeter response
is treated in a realistic way using a parametrization of the electromagnetic and
hadronic shower deposits obtained from a full GEANT simulation~\cite{bramhs} and
including a cluster finding algorithm. Pattern recognition is emulated
by means of a complete cross-reference table between generated particles and
detector response. The output of the program consists in a list of reconstructed
objects: electrons, gammas, muons, charged and neutral hadrons and unresolved
calorimeter clusters, as a result of an idealized Energy Flow algorithm 
incorporating track-cluster matching.

\subsection{B-tagging}
\label{btagging}
 Jets coming from $b$ and $c$-quark decays are tagged based on the non-zero lifetime of these
quarks, using the Vertex Detector (VDET). In this study we have assumed the performance of a 
CCD VDET in a 1 cm radius beampipe.

 In order to look for this lifetime signal, we have chosen to use the 3D impact parameter (IP) 
of each charged track (distance of closest approach between the 
track and the $b$ production point). Since the statistical resolution of the IP
varies strongly from one track to another, we use the estimated statistical significance of the 
measured IP to define our tag. The b-tagging algorithm is kept simple so that the success of the analysis
does not depend on detector details. More efficient algorithms can be developed by making use
of multivariate techniques, such as Neural Networks.

\begin{figure}[h]
\begin{center}
\mbox{
\epsfig{file=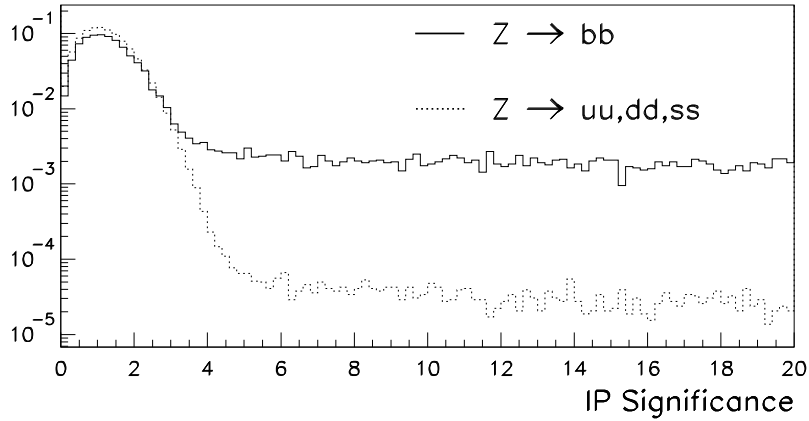,width=13cm}
}
\end{center}
\caption{\label{btag_3dip}
\protect\footnotesize
Track 3D impact parameter significance for $Z$ hadronic events at $\sqrt{s}$=100 GeV.}
\end{figure}

 In Fig.~\ref{btag_3dip}, the IP significance distributions for different $Z$ hadronic
decays are compared. The lifetime signature can be clearly seen for $Z \to b \bar{b}$ in the 
positive tail.

 We will use the IP significance distribution for non-lifetime tracks (those originated from 
$Z \to u\bar{u},d\bar{d},s\bar{s}$)
to define, for each track, a 
probability ``to be consistent with being originated from the primary vertex''. This information
can then be combined to get also a probability per jet or per event~\cite{aleph}. 

 In order to test the performance of such b-tagging, we have 
estimated its efficiency and purity for a given cut in the jet probability. To do so, the Monte Carlo
generated quarks are assigned to the reconstructed jets by a matching algorithm which
associates those quark-jet pairs with minimum invariant mass, starting from the most energetic quark.
Now, an efficiency $\epsilon_b$ and purity $\rho_b$ can be defined as: 
\begin{center}
\bea
\epsilon_b = \frac{n_{b,corr}}{n_b}, \hspace{0.5cm} {\rm and} \hspace{0.5cm} \rho_b = \frac{n_{b,corr}}{n_{b,tag}} \nonumber
\eea
\end{center}
\noindent where $n_{b,corr}$ is the number of b-jets correctly tagged,
$n_{b,tag}$ is the total number of tagged b-jets,
and ${n_b}$ is the actual number of b-jets in the event. In the way the purity is defined here, 
it measures the missidentification
probability of jets in a given process.

\begin{figure}[h]
\begin{center}
\mbox{
\epsfig{file=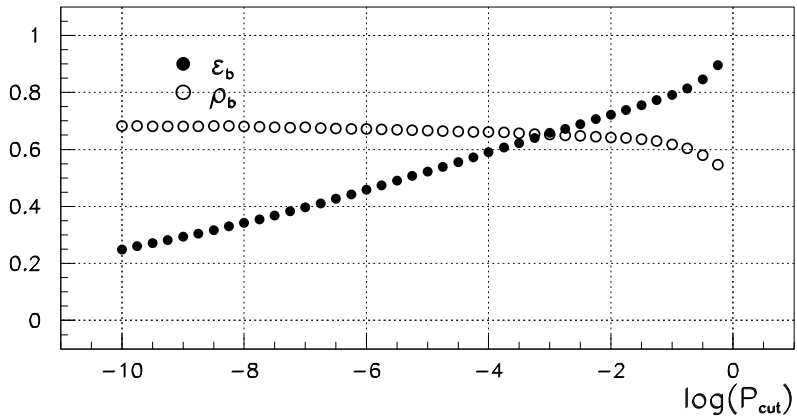,width=13cm}
}
\end{center}
\caption{\label{btag_effpur}
\protect\footnotesize
b-tagging efficiency and purity for $t\bar{t}H \to q\bar{q}q\bar{q}b\bar{b}b\bar{b}$ events 
as a function of the lifetime probability cut.}
\end{figure}

 The b-tagging efficiencies and purities shown in Fig.~\ref{btag_effpur} as a function of the jet 
probability cut,
correspond to a sample of signal events where the $W$ leptonic decays have been switched off
and the Higgs has been forced to decay into $b\bar{b}$ ($t\bar{t}H \to q\bar{q}q\bar{q}b\bar{b}b\bar{b}$). 
It should be noted that, due to the high multiplicity of such events, 
the probability that
the jet clustering algorithm assigns ``lifetime tracks'' (coming from the b-jets) to a
light-quark ($uds$) jet is large.
This leads to rather low values of the b-tagging efficiency such as $\epsilon_b \sim 80\%$ for a purity of $\rho_b \sim 60\%$. 
In order to quantify the performance degradation of the b-tagging as the event multiplicity
increases,
this algorithm has been tested with $ZZ$ events, where one of the $Z$ bosons
has been forced to decay into $b$ quarks and the other into light-quarks. The achieved 
b-tagging efficiency in this case is $\epsilon_b \sim 80\%$ for a purity of $\rho_b \sim 80\%$.

Given the low $\epsilon_b$ values for $t\bar{t}H$ events,
the efficiency to tag correctly the 
four b-jets of a signal event by fixing a probability cut
will be in general very small.
In order not to reduce drastically the signal efficiency, we will not 
use the number 
of found b-jets for a certain lifetime probability as a selection cut. Instead, for every event,
we will define as 
b-jets those four with the lowest probability (to originate from the primary vertex). 
Applied to purely hadronic signal events, for example, this algorithm will tag 
the four correct b-jets in $\sim37\%$ of the cases, and at least three of them in $\sim88\%$
of the cases.    

\section{Experimental Analysis}\label{analysis}
The experimental analysis is performed assuming a 
total integrated luminosity of 1000 fb$^{-1}$, which
can be collected in around 3 years of running at ${\cal L} = 10^{34}$
cm$^{-2}$s$^{-1}$.

Both semileptonic and fully hadronic decay channels have been considered.
In spite of the apparently clean signature of both channels ($\geq 6$ jets
in the final state, out of which $\geq 4$ are b-jets, multi-jet invariant 
mass constraints, etc), the measurement has many difficulties, among which:
\begin{itemize}
\item tiny signal ($\simeq 2.6$ fb) in front of backgrounds about 3 orders
of magnitude larger: in Table~\ref{tb:sg}, the total cross-section for the
signal and different backgrounds considered is listed together with the number of
generated events.
\item limitations of jet-clustering algorithms in properly reconstructing
multi-jets in the final state due to hard gluon radiation, jet mixing, etc,
\item degradation of b-tagging performance due to hard gluon radiation and jet mixing.
\end{itemize}

\begin{table}[htpb]
\begin{center}
\begin{tabular}{lrc}
\hline\hline 
{\rm Process} & {$\sigma$ \rm (fb)} & {Generated events} \\
\hline
$t\bar{t}H$ & 2.57 & 100k \\
\hline
$q\bar{q} \rm \:(5\:flav.)$ & 1557.7 & 1M\\
$t\bar{t}$ & 303.08 & 1M\\
$t\bar{t}Z$ & 4.57 & 100k\\
$W^+W^-$ & 4402.7 & 3.8M\\
$ZZ$ & 308.9 & 300k\\
$HZ\:(Z \nrightarrow t\bar{t})$ & 24.8 & 100k\\
\hline\hline
\end{tabular}
\caption{\label{tb:sg}\protect\footnotesize Total cross-section for signal and the different backgrounds considered at $\sqrt{s}$ = 800 GeV. Initial state radiation and 
beamstrahlung have been included. Also listed is the number of generated
events for every process.}
\end{center}
\end{table}   

Due to the extremely small signal-to-background (S/B) ratio, 
the philosophy of the analysis in both decay channels
will be to start by applying a standard cuts preselection in order to remove as much background 
as possible while keeping a high efficiency for the signal. Then, in order to further improve
the statistical sensitivity, a multivariate analysis will be performed. At this stage our problem 
will be how to make an optimal use of the statistical information from a set of $N$ distributions
discriminating between signal and background. It can be proven~\cite{optproj} 
that it is possible to make
an optimal projection from this input $N$-dimensional space to a 1-dimensional 
space\footnote{In the general case
of $m$ existing classes to be discriminated, the optimal projection is performed in 
a space ($m-1$)-dimensional. In our problem, all backgrounds are considered inclusively and
$m=2$, thus the optimal projection being 1-dimensional.}:
\begin{description}
\item{a)} without loss of sensitivity on the classes proportions and
\item{b)} with a probabilistic interpretation (in terms of the a-posteriori Bayesian probability
of being of signal type).
\end{description}
This projection can be performed by using Neural Network (NN) techniques, which have become
increasingly popular in High Energy Physics in the last few years.

Even after selection, a S/B ratio much larger than 1 can only be obtained at the expense
of a rather low signal efficiency. We have considered that the uncertainty on the 
background normalization after selection is going to be the dominant contribution to the
systematic error.
The main concern is how well parton shower can reproduce the tails in the distributions for those
non-interfering background processes. In order to be just limited by that, it is important to
have available event generators with 8 fermions in the final state, thus properly
accounting for all interfering backgrounds. 
The main background process after selection is $t\bar{t}$.
Therefore, theoretical calculations up to ${\cal O}(\alpha_s^2)$ would be needed by the time this
measurement is performed. 

For a given systematic uncertainty in the background normalization, it is possible to 
adjust the selection signal efficiency in order to optimize the
total uncertainty in the Yukawa coupling. We have set as a goal a 5\% systematic 
uncertainty in the background normalization in order not 
to dominate the measurement. Then, we have optimized the selection assuming this
uncertainty.

\subsection{Semileptonic Channel}
The semileptonic decay channel, with a branching ratio of 43.9\%,  
constitutes the golden channel in terms of high statistics and clean signature
as compared to the fully hadronic decay channel, where 8 jets have to be reconstructed in
the final state. The final state is:
\begin{eqnarray}
e^+e^- \to t\bar{t}H \to q\bar{q}b\: \ell^{\pm}\nu_{\ell}\bar{b}\:b\bar{b}, \nonumber
\end{eqnarray}
\noindent the experimental signature being then: 4 b-jets + 2 light-quark jets + $\ell^{\pm}$ + $E_{miss}$. 
Hence, the high $p_T$ and isolated lepton can be used for triggering purposes and
might provide clear separation in the offline selection as well. The four-momentum imbalance 
due to the neutrino presence will also represent a discriminant variable as long as
the final detector has a good hermeticity.
Finally, the high content in b-jets will also be exploited as a powerful
discriminant variable by using the vertex detector.

\begin{figure}[h]
\begin{center}
\mbox{
\epsfig{file=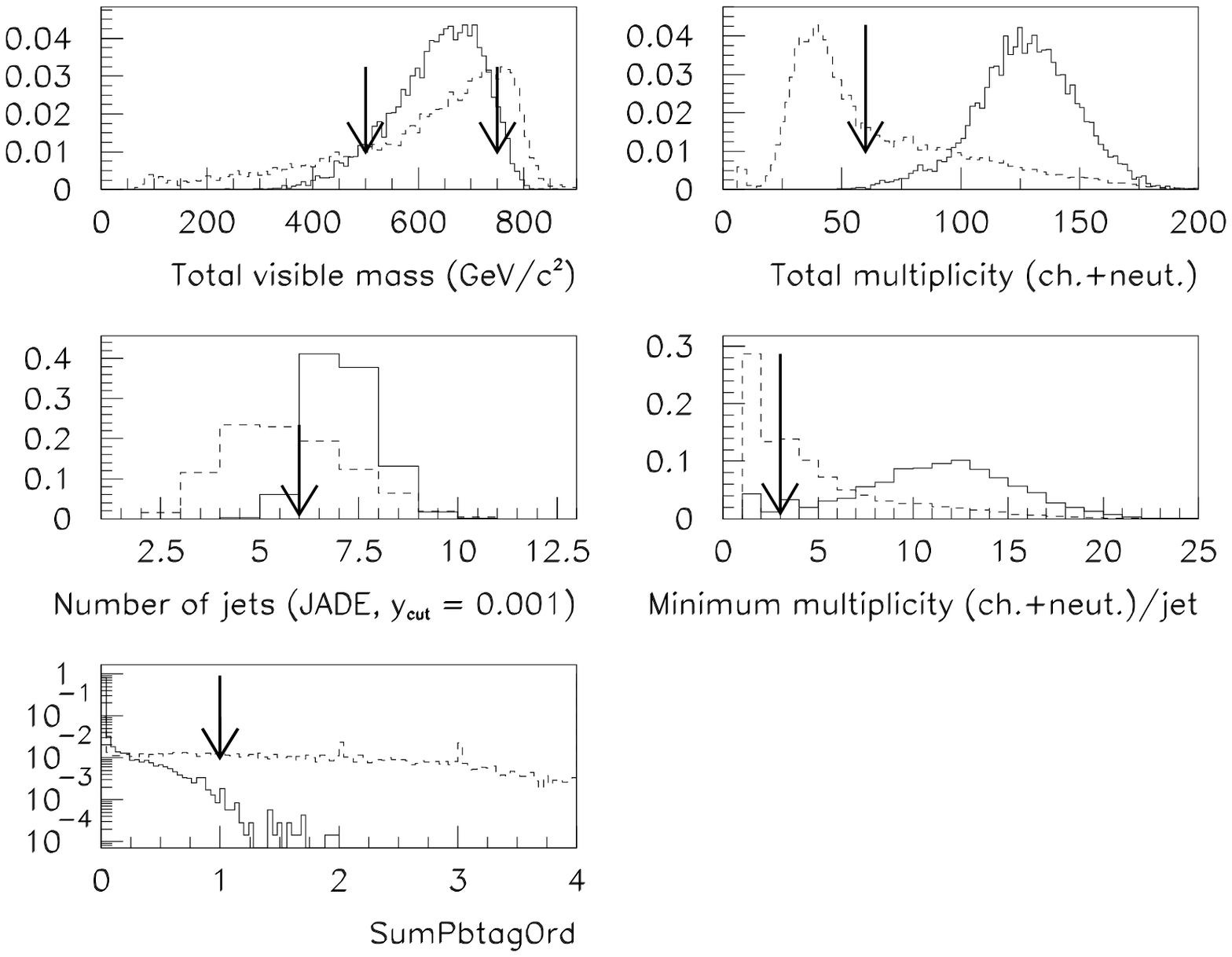,width=13cm}
}
\end{center}
\caption{\label{sl_presel_notopo}\protect\footnotesize
Preselection variables for the semileptonic channel (I).
Signal (solid) and background (dashed) have been normalized to the same
number of events.
The background histograms have been
built by adding all the different backgrounds contributions weighted according to their
relative cross-sections.}
\end{figure}

\begin{figure}[h]
\begin{center}
\mbox{
\epsfig{file=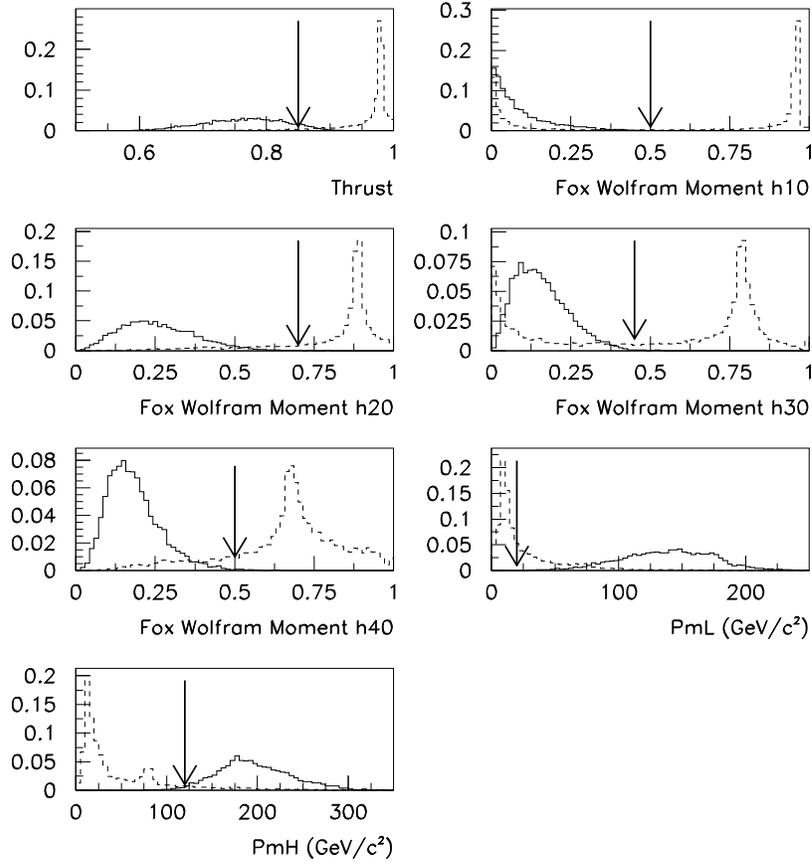,width=13cm}
}
\end{center}
\caption{\label{sl_presel_topo}
\protect\footnotesize
Preselection variables for the semileptonic channel (II).
Signal (solid) and background (dashed) have been normalized to the same
number of events.
The background histograms have been
built by adding only the $W^+W^-$ and $q\bar{q}$ contributions weighted according to their
relative cross-sections. In this way the differences between the signal and 
the most topologically different backgrounds are clearly visualized.}
\end{figure}

        As already stated, the philosophy of the selection has been to start with 
a series of preselection cuts addressed to remove as much background events as possible
while keeping high efficiency for the signal. 
The preselection variables are compared for signal and background in Figs.~\ref{sl_presel_notopo}
and~\ref{sl_presel_topo}, 
along with the cuts applied.
The selected events are required to have a visible mass larger than 500 GeV but lower
than 800 GeV, more than 60 energy flow (EF) objects reconstructed and at least 6 jets 
reconstructed with the JADE~\cite{JADE} jet-clustering algorithm for a resolution 
parameter $y_{cut} = 10^{-3}$.
Then a series of cuts on topological variables such as the thrust 
and the normalized Fox Wolfram moments of the event are applied. 
These cuts are mainly addressed to drastically reduce the contamination of high cross-section backgrounds such as $W^+W^-$ or radiative $q\bar{q}$,
which tend to be much less spherical than signal events 
(due to the boost) and have a large value for these variables (see Fig.~\ref{sl_presel_topo}).
Also useful are the so-called high and low jet masses of the event, $PmH$ and $PmL$, respectively. 
The event is divided in two hemispheres and particles are assigned to either hemisphere 
in order to minimize the quadratic sum of the two hemispheres invariant masses. 
For processes with two resonances (such as $W^+W^-$ or $ZH$), 
these distributions tend to show resonant structures around the true invariant masses.

        At this stage, and in order to reconstruct the $t\bar{t}H$ semileptonic decay signature, 
an energetic and isolated lepton has to be identified.
This lepton candidate has been chosen as the charged track which maximizes 
$E_\ell(1-cos\theta_{\ell j})$, where 
$E_\ell$ is the track energy and $\theta_{\ell j}$ the angle of such track with 
the closest of the 6 jets 
to which the remaining EF objects have been forced by using JADE.
The efficiency of such algorithm to find the correct lepton (the one from the $W$ decay) for 
$t\bar{t}H$ semileptonic events has been determined to be $\sim 98\%$.

        Once the event has been clustered into 6 jets plus 1 lepton and 
after rejecting those events having jets with less than 3 EF objects,
a probability to contain no-lifetime is assigned to each jet as
described in Sect.~\ref{btagging}. This allows us to build a powerful variable which
reflects the lifetime content of the event by adding up the probabilities of the 
four most b-like jets (those with lower non-lifetime probability).  

        In Table~\ref{sl_Pres_Effs} the preselection efficiencies for the signal and the 
different backgrounds are displayed. The situation is such that the overall effective cross-section
for the background is 17.60 fb, while for the signal is only 0.61 fb.
This translates into such a poor sample purity ($\rho \sim 3.3 \%$),
that any uncertainty in the background normalization completely erases 
any significance in the signal. 

\begin{table}[htpb]
\begin{center}
\begin{tabular}{lrc}
\hline\hline 
{\rm Process} & {$\epsilon$ (\%)} & $\sigma_{eff}$ {\rm (fb)} \\
\hline
$t\bar{t}H \to 6q\ell\nu$ & 54.08 & 0.61 \\
\hline
$t\bar{t}H \to 8q+4q\ell\nu\ell\nu$ & 18.80 & 0.27 \\
$q\bar{q} \rm (5\:flav.)$ & 0.102 & 1.18\\
$t\bar{t}$ & 3.621 & 10.97\\
$t\bar{t}Z$ & 21.08 & 0.96\\
$W^+W^-$ & 0.092 & 4.05\\
$ZZ$ & 0.053 & 0.16\\
\hline
{\em Total Bckg} & & 17.59 \\
\hline\hline
\end{tabular}
\caption{\label{sl_Pres_Effs}\protect\footnotesize 
Semileptonic channel preselection efficiencies and effective cross-sections.}
\end{center}
\end{table}   

        However, there are several variables which, after the 
preselection, still have some discriminating power. We will use a NN in order to
project in an optimal way this N-dimensional information into a 1-dimensional variable.

\begin{figure}[h]
\begin{center}
\mbox{
\epsfig{file=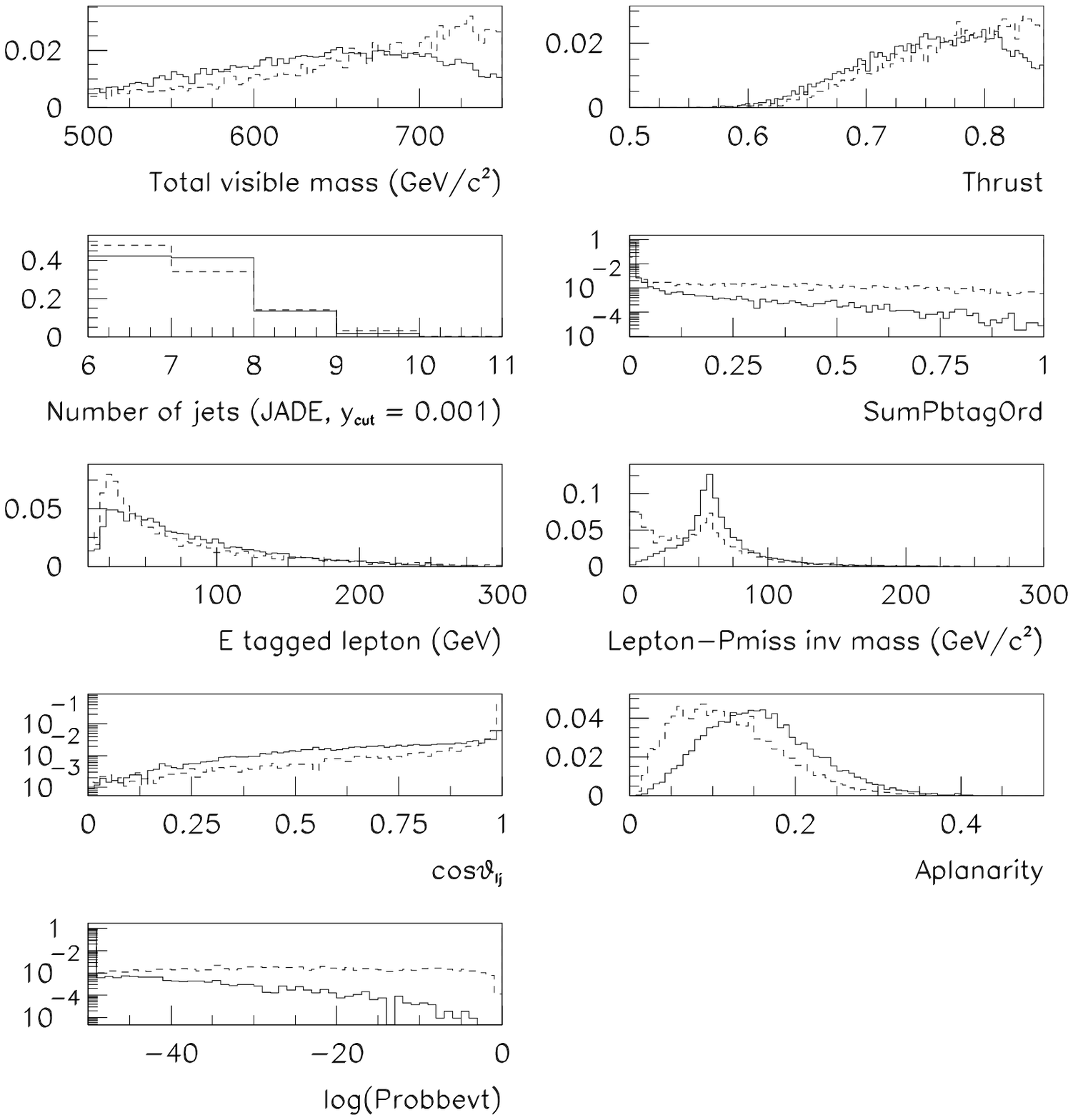,width=13cm}
}
\end{center}
\caption{\label{sl_nn_vars}
\protect\footnotesize
Selection Neural Network variables for the semileptonic channel.
Signal (solid) and background (dashed) have been normalized to the same
number of events.
The background histograms have been
built by adding all the different backgrounds contributions weighted according to their
relative cross-sections.}
\end{figure}

        Nine variables showing high discrimination at the preselection level 
(see Fig.~\ref{sl_nn_vars}) have been chosen to train a NN to separate the signal
from the background. Three of these variables use the information of the leptonic decay of one
of the $W$ bosons (lepton energy, angle between lepton and closest jet and invariant mass between the
lepton and the missing momentum), 
two use the b-jets content of the event (global no-lifetime probability for the event and
sum of the no-lifetime probabilities of the four most b-like jets) and four more topological
variables (thrust, aplanarity, number of jets clusterized 
with JADE for $y_{cut}=10^{-3}$ and the total visible mass).

        After training the NN, the weights distribution for the 
different neurons allows to determine the discriminant power of each of the
9 input variables (see Table~\ref{sl_nn_discpower}). It can be
seen that the most discriminating variables are those containing b-tagging
information, followed by the identified lepton related variables.

\begin{table}[htpb]
\begin{center}
\begin{tabular}{lc}
\hline\hline 
{\rm Variable} & {\rm Discriminant Power (\%)} \\
\hline
${\rm M_{vis}}$               & 7.2 \\
${\rm N_{jets} (JADE)}$         & 9.0 \\ 
${\rm \sum_{i=1,4} P_{btag}^{jet\:i}}$ & 21.4\\ 
${\rm Log(P_{btag}^{evt})}$         & 14.4\\ 
Thrust                  & 8.5  \\ 
Aplanarity              & 7.8  \\ 
${\rm E_\ell}$                   & 11.8 \\ 
${\rm M_{\ell\nu}}$          & 10.8 \\ 
${\rm cos\theta_{\ell j}}$        & 9.1  \\ 
\hline\hline
\end{tabular}
\caption{\label{sl_nn_discpower}\protect\footnotesize 
Discriminant power of each of the 9 input variables of the semileptonic selection Neural Network.} 
\end{center}
\end{table}   
 
\begin{figure}[h]
\begin{center}
\mbox{
\epsfig{file=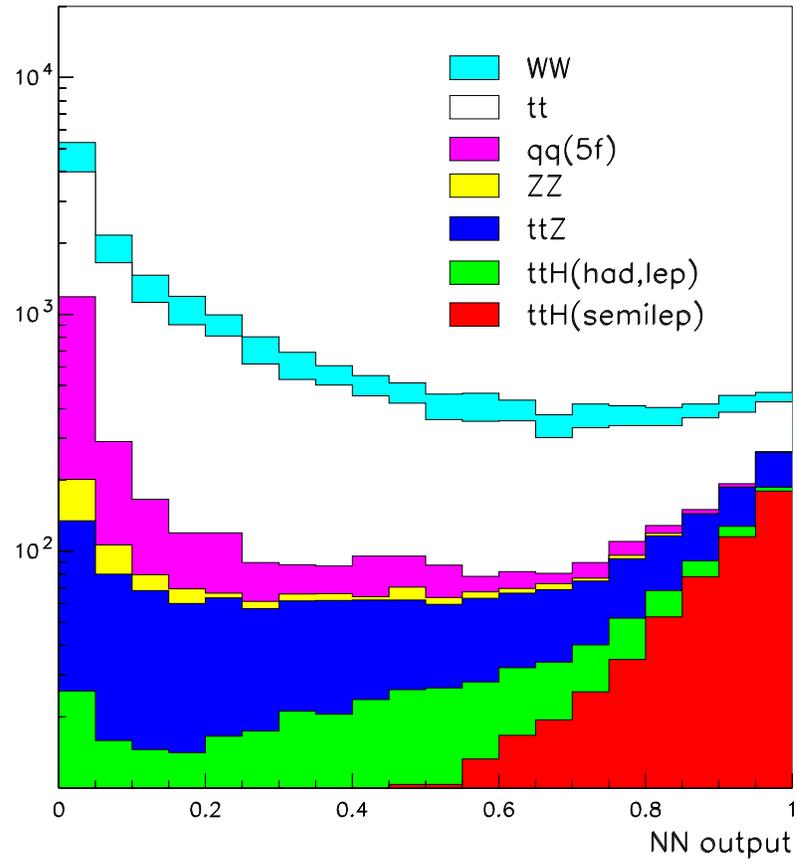,width=13cm}
}
\end{center}
\caption{\label{sl_nno}
\protect\footnotesize
Semileptonic Selection Neural Network output. The different contributions are normalized
to the same luminosity (1000 fb$^{-1}$).}
\end{figure}

        The NN output distributions for the signal and the 
different backgrounds are compared in Fig.~\ref{sl_nno}. The histograms have been 
normalized to the same integrated luminosity (1000 fb$^{-1}$) to show how, even
after having used the information in the 9 variables, the S/B
ratio is smaller than one in all the bins. 

        Applying a cut in the NN output variable allows us to perform
an optimal test of decision. A measurement of the $t\bar{t}H$ cross-section, and hence of
the top Yukawa coupling, can then be done assuming the knowledge of the expected number of 
background events. However, is necessary to take into account the systematic uncertainty
in the background normalization which, as already mentioned in Sect.~\ref{analysis}, has been assumed
here to be at the 5\% level.

\begin{figure}[h]
\begin{center}
\mbox{
\epsfig{file=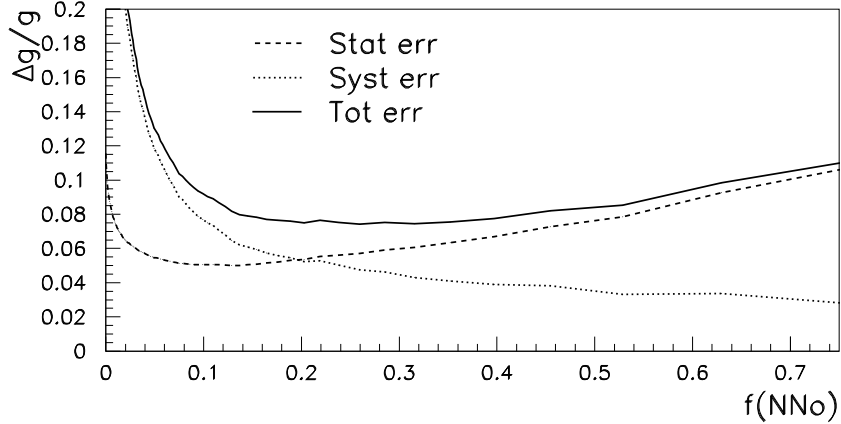,width=13cm}
}
\end{center}
\caption{\label{sl_nncutscan}
\protect\footnotesize
Evolution of the error in the top Yukawa coupling as a function of the NN output cut 
(semileptonic channel).}
\end{figure}

        In Fig.~\ref{sl_nncutscan} the evolution of both the statistical and systematic uncertainties 
(as defined above) is plotted as the cut in the NN output is varied. Actually, the
horizontal scale is not directly the NN output, but the a-posteriori Bayesian probability of being of
signal-type for the expected proportions of signal and background, 
which can be computed as a function of the NN output.
It can be seen how, by cutting harder on the NN output, the systematic error coming from the
background normalization can be kept under control since a higher sample purity is achieved.
The optimal cut is found within the plateau in the total error for the minimum possible 
systematic uncertainty.

        In Table~\ref{sl_errors}, the statistical and systematic uncertainties are compared
in different steps, illustrating the usefulness of the multivariate analysis.
As already mentioned, the sample purity after preselection is so poor that, not only the
statistical error degrades because of statistical fluctuations in the background, but the
systematic error (see Eq.~\ref{gttherrors}) 
introduced as a result of 5\% uncertainty in the background normalization
leads effectively to the absence of a measurement. The situation is dramatically improved
when using a multivariate analysis and the top-Higgs Yukawa coupling is determined from a
fit to the output neuron distribution. The systematic uncertainty can be further controlled
(at the expense of slightly increasing the statistical uncertainty) by choosing a suitable cut
in the NN output distribution.

\begin{table}[htpb]
\begin{center}
\begin{tabular}{lccccc}
\hline\hline 
 & $\epsilon\:(\%)$ & {\rm S/B} & {\rm Stat. error (\%)} & {\rm Syst. error (\%)} & {\rm Total error (\%)}    \\
\hline
{\rm After Preselection} & $54.1$ & $0.03$ & $11.5$ & $63.1$ & $64.1$ \\
{\rm Fit to NNSel (50 bins)} & $''$ & $''$ & $4.4$ & $9.1$ & $10.1$ \\
{\rm NNSel$>$0.9 (optimal cut)} & $27.1$ & $0.51$ & $5.1$ & $3.8$ & $6.3$ \\ 
\hline\hline
\end{tabular}
\caption{\label{sl_errors}\protect\footnotesize 
Statistical and systematic uncertainties in the top-Higgs Yukawa coupling
from the semileptonic channel. An integrated luminosity of 1000 fb$^{-1}$ has
been assumed.}
\end{center}
\end{table}   

\subsection{Hadronic Channel}
The fully hadronic decay channel, with a branching ratio of 45.6\%,
benefits from the high statistics but is the most difficult one in terms
of signal to background discrimination, experimental reconstruction 
and the one potentially more affected by systematic uncertainties. The final state is:
\begin{eqnarray}
e^+e^- \to t\bar{t}H \to q\bar{q}b\:q\bar{q}\bar{b}\:b\bar{b}, \nonumber
\end{eqnarray}
\noindent the experimental signature being quite challenging: 8 jets in the final state, out of which 
4 are b-jets. This imposes quite stringent requirements in the vertex detector.
Due to limitations in the jet clustering algorithms, the ability to properly reconstruct
8 jets is not dominated by the detector performance. In this sense, this channel
does not impose strict requirements for what the tracker momentum resolution and
calorimeter energy resolution and granularity is concerned.

Potential backgrounds are genuine multi-jet processes like $t\bar{t}$ and
$t\bar{t}Z$. In fact, the large cross-section of $t\bar{t}$, together with hard gluon
emission easily emulating 8 jets make of this process the most important background after selection. 
Of particular concern is $t\bar{t}g^*$, with the gluon splitting into a $b\bar{b}$ pair, since it 
easily emulates 4 b-jets in the final state, even though the invariant mass of the $b\bar{b}$ pair
from the gluon splitting peaks at low values.
Since the assumed Higgs mass is relatively close to the $Z$ mass,
$t\bar{t}Z$ constitutes an almost irreducible background. The main reason is that
the invariant mass resolution of the $b\bar{b}$ system becomes seriously degraded 
due to particle mixing between jets in such a populated environment and to energy losses
in neutrino emission from the B-meson cascade decays. 
On the other hand and again due to the large cross-section and limitations in the b-tagging in such a busy
environment,  hadronic final
states such as $W^+W^- \to q_1\bar{q}_2q_3\bar{q}_4g^*$, where the gluon splits into a $b\bar{b}$ pair
will lead to the necessity of a stronger selection and therefore reduced statistical
sensitivity on the Yukawa coupling.

\begin{figure}[p]
\begin{center}
\mbox{
\epsfig{file=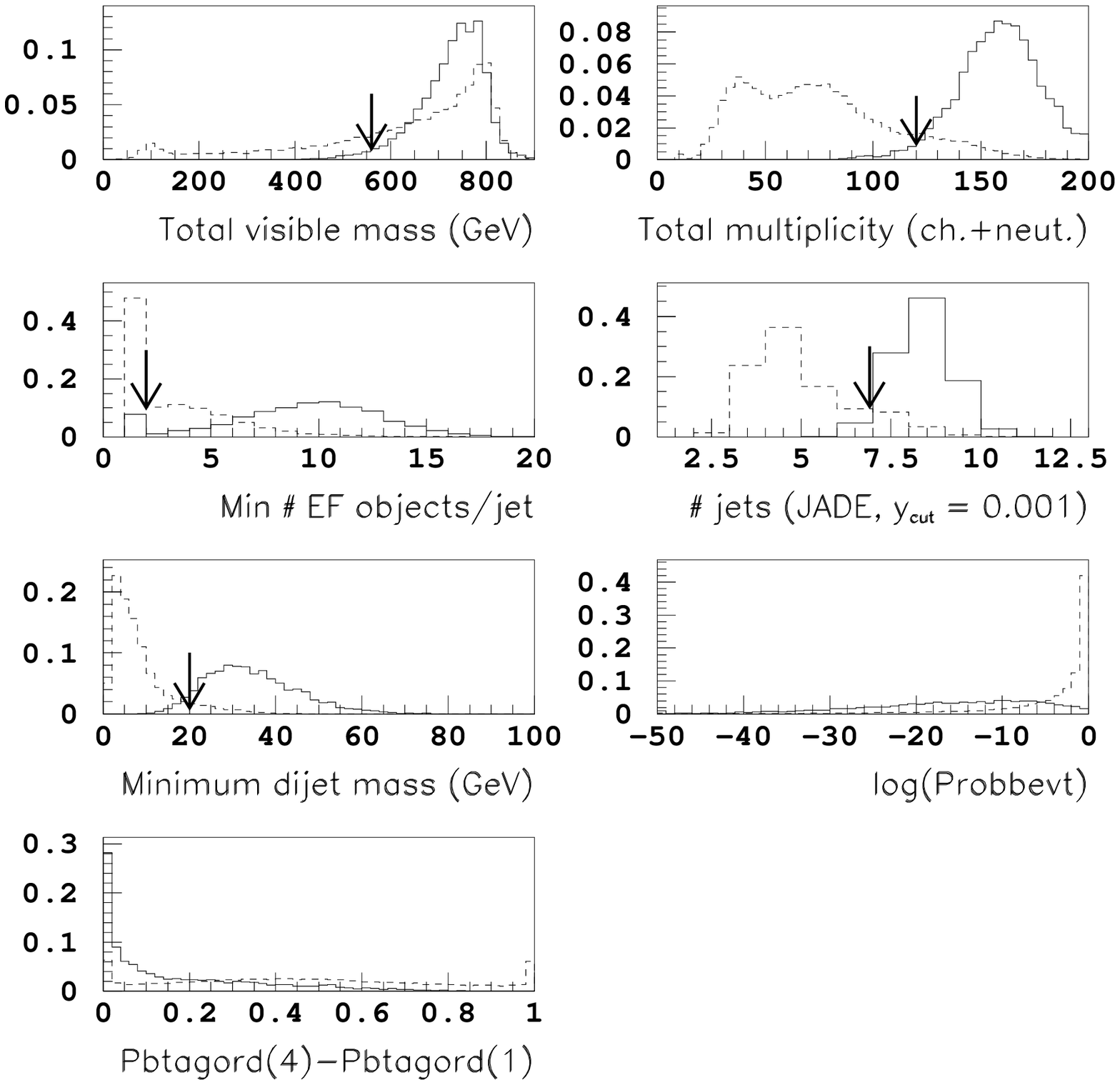,width=13cm}
}
\end{center}
\caption{\label{had_presel_cuts}
\protect\footnotesize
Preselection variables for the hadronic decay channel.
Signal (solid) and background (dashed) have been normalized to the same
number of events.
The background prediction has been computed by adding all the different 
background contributions weighted according to their relative cross-section.}
\end{figure}

Like for the semileptonic channel, a standard cuts preselection is applied in order
to remove as much background as possible before the multivariate analysis.
The selected events are required to have a visible mass in excess of 70\%$\sqrt{s}$
(that is 560 GeV), more than 120 EF objects reconstructed and
at least 7 jets reconstructed with the JADE algorithm for a resolution parameter
$y_{cut} = 10^{-3}$.
Then the event is forced to have exactly 8 jets reconstructed by using JADE.
Further preselection cuts require a minimum of 2 EF objects per jet, a minimum
di-jet invariant mass of 20 GeV and the thrust larger to 0.85. 
The preselection variables are compared
for signal and background in Fig.~\ref{had_presel_cuts}, along with the
cuts applied. The preselection efficiencies and effective
cross-section for the different processes considered are listed in 
Table~\ref{had_Pres_Effs}.

\begin{table}[htpb]
\begin{center}
\begin{tabular}{lrc}
\hline\hline 
{\rm Process} & {$\epsilon$ (\%)} & $\sigma_{eff}$ {\rm (fb)} \\
\hline
$t\bar{t}H \to 8q$ & 77.06 & 0.90 \\
\hline
$t\bar{t}H \to 6q\ell\nu+4q\ell\nu\ell\nu$ & 9.63 & 0.14 \\
$q\bar{q} \rm \:(5\:flav.)$ & 0.378 & 4.38\\
$t\bar{t}$ & 5.02 & 15.22\\
$t\bar{t}Z$ & 27.35 & 1.25\\
$W^+W^-$ & 0.185 & 8.14\\
$ZZ$ & 0.139 & 0.43\\
\hline
{\em Total Bckg} & & 29.55 \\
\hline\hline
\end{tabular}
\caption{\label{had_Pres_Effs}\protect\footnotesize Hadronic channel preselection efficiencies and effective cross-sections.}
\end{center}
\end{table}   

After preselection, the efficiency for signal is reduced to 77\% and the sample
purity is only $\simeq 3.0\%$.

Note the high remaining cross-section for $q\bar{q}$ and $W^+W^-$ despite the cuts
applied, e.g. in minimum di-jet invariant mass.
The main responsible is hard gluon radiation. In Table~\ref{hardgluon}, the average number
of gluons with $P_g>20$ GeV/c, average gluon momentum and angle of the gluon with respect to the
parent quark are listed for both processes, before and after preselection.
Indeed, multi-gluon radiation (as simulated by parton shower)
leads to genuine 8-jet final state even for processes like $q\bar{q}$ (5 flav.)/$W^+W^-$
with only 2/4 initial partons. Owing to the large cross-section of these processes,
they constitute non-negligible backgrounds which have to be taken into account in
any ``realistic simulation''.

\begin{table}[htpb]
\begin{center}
\begin{tabular}{lrcc}
\hline\hline 
{\rm Process} & $\rm <N_g>$ & $\rm <P_g>\:(GeV/c)$ & $\rm <\theta_{gq}>\:(deg.)$ \\
\hline
$q\bar{q} \rm \:(5\:flav.)$ & $4.2(6.9)$ & $64.6(50.0)$ & $16.0(44.4)$\\
$W^+W^-$ & $3.8(4.0)$ & $61.6(45.4)$ & $6.3(22.7)$\\
\hline\hline
\end{tabular}
\caption{\label{hardgluon}\protect\footnotesize
Hard gluon ($P_g>20$ GeV/c) radiation as predicted by parton shower at $\sqrt{s}=800$ GeV
before and after (between parenthesis) preselection.
}
\end{center}
\end{table}   

\begin{figure}[htpb]
\begin{center}
\mbox{
\epsfig{file=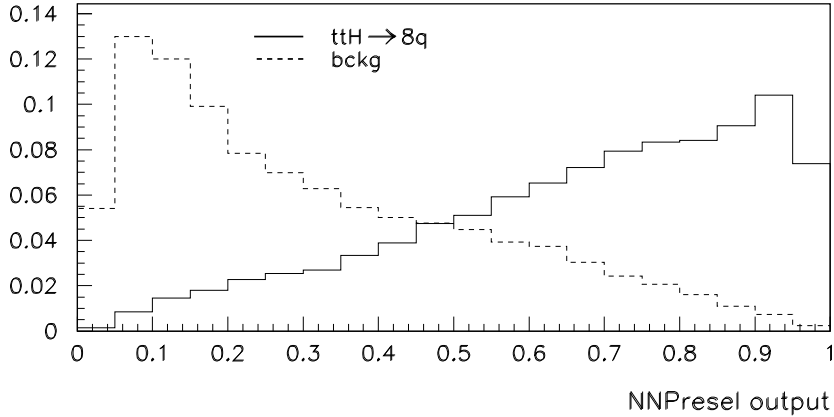,width=13cm}
}
\end{center}
\caption{\label{nhad_nout_presel}
\protect\footnotesize
Hadronic preselection Neural Network output.
Signal (solid) and background (dashed) have been normalized to the same
number of events.}
\end{figure}

As it can be observed in Fig.~\ref{had_presel_cuts}, the preselection variables
after cuts still have discriminating power between signal and background. In order to
optimally use these variables, they are further used together with two more
variables to train a Preselection Neural Network. The two variables added (shown as the two
last variables in Fig.~\ref{had_presel_cuts}) provide information
about the lifetime content of the event: the logarithm of the event probability
to contain no-lifetime and the difference between the probability of the fourth
jet and the first jet (sorted from the most b-like to the least b-like).
In Fig.~\ref{nhad_nout_presel} it is shown the Preselection Neural Network output, after training,
for both signal and background. No cut is applied in this distribution, but it is rather used as a
discriminant variable.

\vspace{-1.0cm}

\newpage 

\begin{figure}[p]
\begin{center}
\mbox{
\epsfig{file=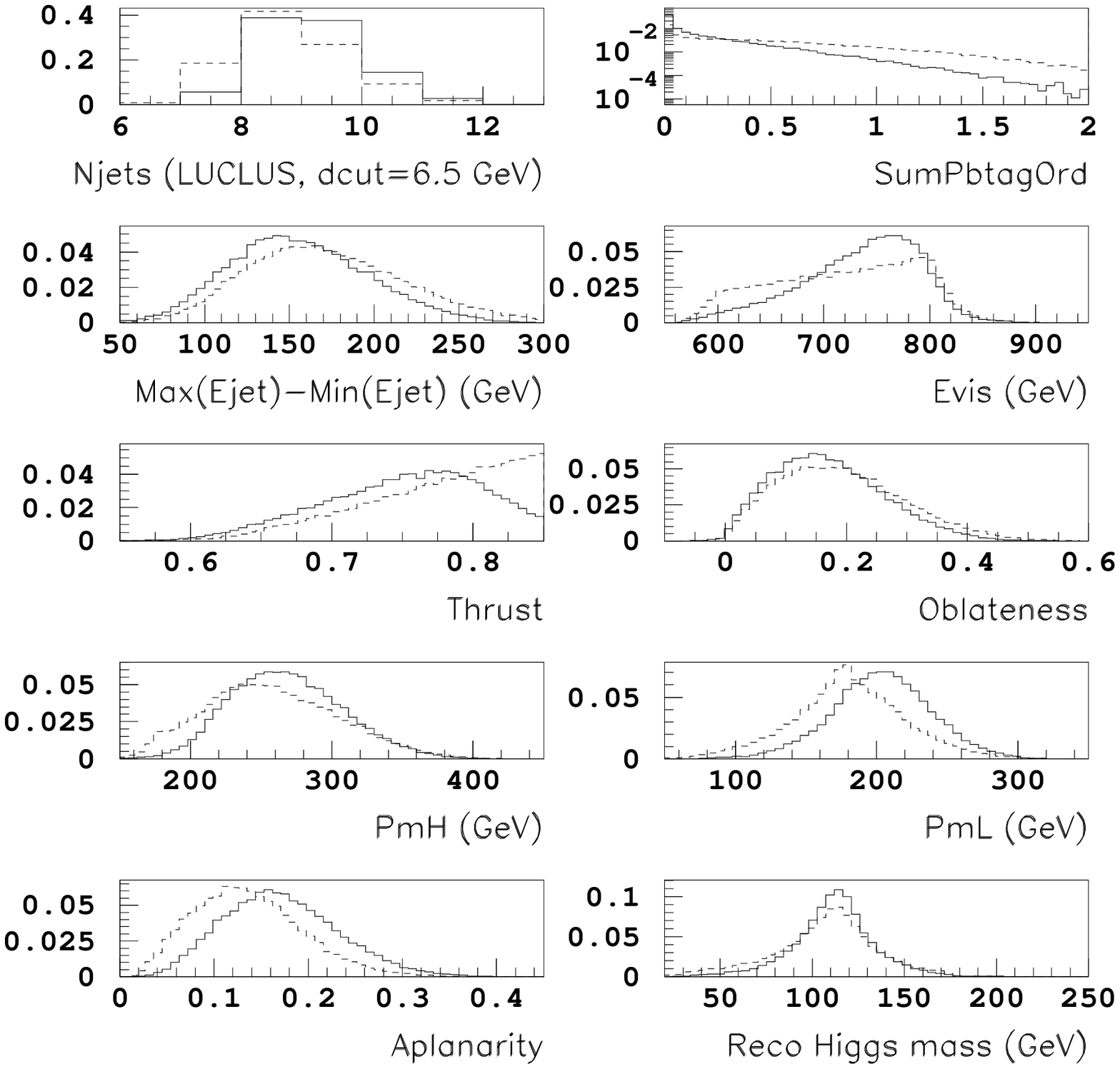,width=13cm}
}
\end{center}
\caption{\label{had_nn_vars}
\protect\footnotesize
Selection Neural Network variables for the hadronic decay channel.
Signal (solid) and background (dashed) have been normalized to the same
number of events.
The background prediction has been computed by adding all the different 
background contributions weighted according to their relative cross-section.}
\end{figure}

There are 10 more variables which are discriminating between signal and background (see Fig.~\ref{had_nn_vars}). Most of them are variables about the global event topology:
\begin{itemize}
\item Njets(LUCLUS, dcut=6.5 GeV): number of jets found with the LUCLUS~\cite{LUCLUS} jet-clustering
      algorithm for a distance measure of 6.5 GeV;
\item PmH and PmL: high and low jet masses of the event, already described;
\item Max(Ejet)-Min(Ejet): difference between maximum and minimum jet energy;
\item Evis: total visible energy of the event;
\item Thrust, Oblateness, Aplanarity,
\end{itemize}
\noindent others contain information about flavor tagging (SumPbtagOrd: sum of the probability to contain no-lifetime for the four most b-like jets) or Higgs mass reconstruction: 
\begin{itemize}
\item Reco Higgs mass: reconstructed Higgs mass. The 4 most b-like jets are assumed to be
      the b-jets, which reduces the number of possible jet assignments to 36.
      The combination which maximizes:
\begin{eqnarray}
{\cal P}(m_{i_1 i_2 i_3}-m_t) \times {\cal P}(m_{i_4 i_5 i_6}-m_t) \times
{\cal P}(m_{i_1 i_2}-m_W) \times {\cal P}(m_{i_4 i_5}-m_W) \times {\cal P}(m_{i_7 i_8}-m_H)\nonumber,
\end{eqnarray} 
\end{itemize}
\noindent is selected. In the above expresion ${\cal P}$ are the probability density functions 
for the correct jet pairing and $m_{ij}$ ($m_{ijk}$) is the invariant mass between
jets $i$ and $j$ ($i$, $j$ and $k$).

\begin{figure}[h]
\begin{center}
\mbox{
\epsfig{file=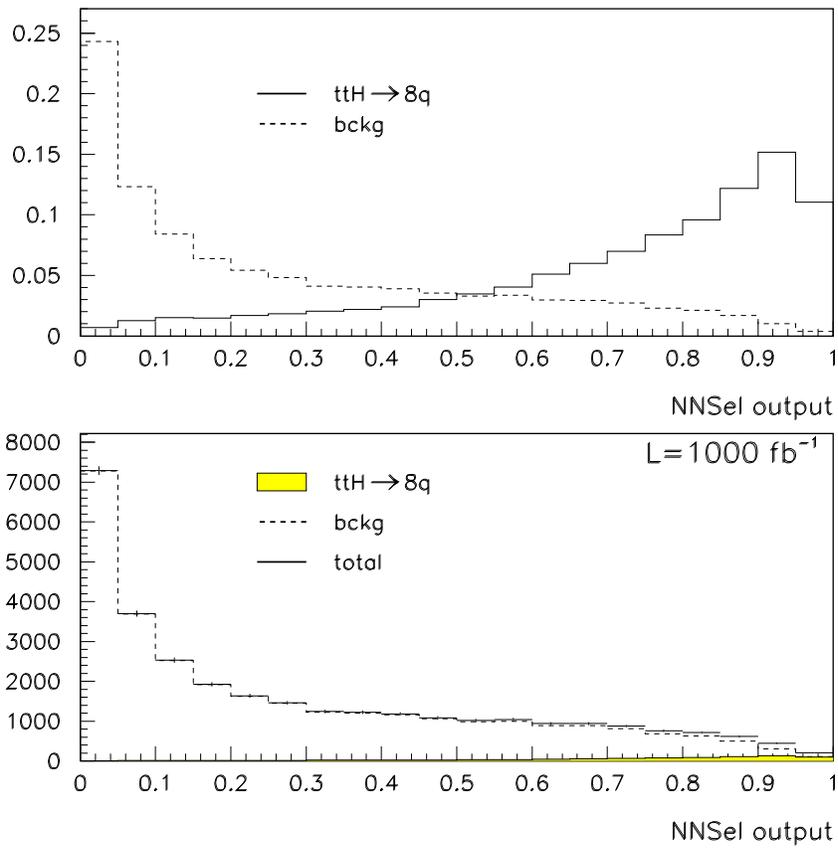,width=13cm}
}
\end{center}
\caption{\label{had_nnout_sel}
\protect\footnotesize
Hadronic selection Neural Network output. Top:
Signal (solid) and background (dashed) have been normalized to the same
number of events. Bottom: comparison of signal (shaded) and background (dashed)
for $L = 1000$ fb$^{-1}$.}
\end{figure}

These variables, together with the Preselection Neural Network output distribution are used
to train a Selection Neural Network. After training, it is found that the most discriminating
variables are the Preselection Neural Network output, reconstructed Higgs mass, thrust and
aplanarity (see Table~\ref{had_selnn_discpower}). The output neuron distribution is
compared for signal and background in Fig.~\ref{had_nnout_sel}.

As for the semileptonic decay channel, 
in Table~\ref{had_errors} the statistical and systematic uncertainty (from 5\%
background normalization uncertainty only) on the top-Higgs Yukawa coupling are
compared in different steps. 

\begin{table}[htpb]
\begin{center}
\begin{tabular}{lc}
\hline\hline 
{\rm Variable} & {\rm Discriminant Power (\%)} \\
\hline
${\rm E_{vis}}$               & 5.9 \\
${\rm max(E^{jet})-min(E^{jet})}$     & 7.6  \\
${\rm N_{jets} (LUCLUS)}$         & 6.9 \\ 
Thrust                  & 13.3  \\ 
Aplanarity              & 11.5  \\
Oblateness              &  5.4  \\
High jet mass           &  7.3  \\
Low jet mass            &  8.2  \\
${\rm \sum_{i=1,4} P_{btag}^{jet\:i}}$ & 6.5\\
${\rm M_H^{reco}}$           & 11.7 \\
${\rm O_{NN}^{presel}}$           & 15.7 \\ 
\hline\hline
\end{tabular}
\caption{\label{had_selnn_discpower}\protect\footnotesize 
Discriminant power of each of the 11 input variables of the hadronic selection Neural Network.} 
\end{center}
\end{table}   

\begin{table}[htpb]
\begin{center}
\begin{tabular}{lccccc}
\hline\hline 
 & $\epsilon\:(\%)$ & {\rm S/B} & {\rm Stat. error (\%)} & {\rm Syst. error (\%)} & {\rm Total error (\%)}    \\
\hline
{\rm After Preselection} & $77.1$ & $0.03$ & $9.8$ & $83.5$ & $83.5$ \\
{\rm Fit to NNSel (50 bins)} & $''$ & $''$ & $4.2$ & $13.7$ & $14.3$ \\
{\rm NNSel$>$0.95 (optimal cut)} & $8.5$ & $0.90$ & $7.3$ & $3.0$ & $7.9$ \\ 
\hline\hline
\end{tabular}
\caption{\label{had_errors}\protect\footnotesize
Statistical and systematic uncertainties in the top-Higgs Yukawa coupling
from the fully hadronic channel. An integrated luminosity of 1000 fb$^{-1}$ has
been assumed.}
\end{center}
\end{table}   

\newpage 

\section{Conclusions}
 The reaction $e^+e^- \to t\bar{t}H$ allows a direct determination of the top-Higgs Yukawa
coupling through its total cross-section measurement. We have studied the feasibility of this
measurement for $M_H = 120$ GeV in an future $e^+e^-$ linear collider operating at 
$\sqrt{s}=800$ GeV and assuming 1000 fb$^{-1}$ of integrated luminosity. 

 The analysis has been performed for both hadronic and semileptonic decay channels,
which constitute almost 90\% of all decays.
For both of them, several sources of background have been considered, 
including not only the interfering ones like
$t\bar{t}Z$ (although interferences haven been neglected)
but also those non-interfering like $q\bar{q}$ or $W^+W^-$ but which have huge
cross-sections as compared to that of the signal.  

 In both cases, a set of variables with high discriminating power has been chosen 
to perform a multivariate analysis in order to use their N-dimensional information
in a way as optimal as possible. For the two studied channels, topological variables 
have been used as well as some others containing b-tagging information. 

 Our final results show the statistical uncertainties that can be achieved 
in each channel for an integrated luminosity of 1000 fb$^{-1}$. 
We have estimated as well the systematic uncertainty that would be associated 
to a 5\% uncertainty in the overall background normalization (even though the main
remaining background is $t\bar{t}$) as a way to 
quantify the importance of an improvement in such theoretical uncertainty by the time the 
measurement might be performed.

 As a final result we can quote the combination of the two channels considering the 
systematic uncertainty fully correlated between them, which leads to a total uncertainty
in the top-Higgs Yukawa coupling of $\simeq 5.5\%$. The statistical uncertainty only would be
$\simeq 4.2\%$.

\section{Acknowledgements}
We are grateful to M. Mart\'\i nez and
P.M. Zerwas for valuable discussions.

\newpage
\newpage


\end{document}